\documentclass[sigconf]{acmart}
\copyrightyear{2026}
\acmYear{2026}
\setcopyright{cc}
\setcctype{by}
\acmConference[SA Technical Communications '26]{SIGGRAPH Asia 2026 Technical Communications}{December 01--04, 2026}{Kuala Lumpur, Malaysia}
\acmBooktitle{SIGGRAPH Asia 2026 Technical Communications (SA Technical Communications '26), December 01--04, 2026, Kuala Lumpur, Malaysia}
\acmDOI{10.1145/3829339.3847814}
\acmISBN{979-8-4007-2841-9/2026/12}

\usepackage{booktabs}
\usepackage{amsmath}
\usepackage{graphicx}

\begin{document}

\title{Local Content-Style Control for Diffusion-based Image Stylization}

\author{Amir Semmo}
\affiliation{%
  \institution{Digital Masterpieces GmbH}
  \city{Potsdam}
  \country{Germany}}
\email{amir.semmo@digitalmasterpieces.com}

\begin{abstract}
Image stylization with latent-diffusion models entangles two independently refined axes: what a region depicts and how it is depicted.
Such pipelines expose only global controls, yet professional retouching demands deliberate, region-specific control.
We lift two conditioning weights already present in a ControlNet\,+\,IP-Adapter stylization pipeline from global scalars to per-location spatial maps, yielding local, per-axis control of content and style in a single generative pass.
Because the two weights act on disjoint pathways, adjusting them independently spans a $2{\times}2$ retouching vocabulary, from free regeneration to identity preservation.
We validate that edits stay confined to the retouched region and that each weight predominantly steers its own axis.
Our approach requires no retraining and drops unchanged into any such pipeline.
\end{abstract}



\begin{teaserfigure}
  \centering
  \includegraphics[width=0.98\textwidth]{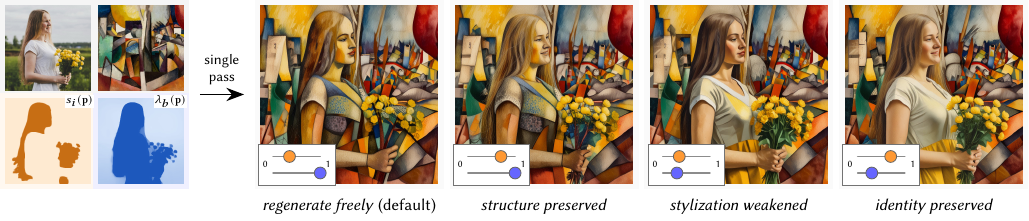}
  \Description{A content photo of a woman holding yellow flowers and a cubist style painting, each with a drawn region map, followed by four stylized results: a freely regenerated cubist rendering; locked content keeping her structure; relaxed style making her appear photographic; and both together preserving her identity amid a cubist background.}
  \caption{Locking content and/or relaxing style spans four local retouching operations.
  Photo by D.~Perevoshchikov on Unsplash.}
  \label{fig:teaser}
\end{teaserfigure}

\maketitle

\section{Introduction}
\label{sec:intro}

\begin{figure*}[t]
  \centering
  \includegraphics[width=0.92\textwidth]{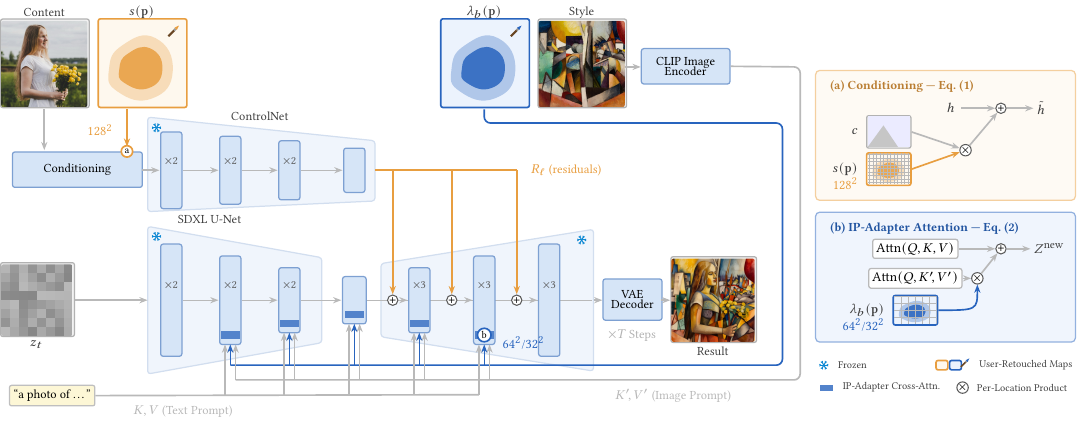}
  \Description{Pipeline diagram of an SDXL U-Net with a ControlNet branch and an IP-Adapter path.
  A user-drawn content map scales the ControlNet control embedding before its residuals enter the U-Net, and a user-drawn style map scales the image-attention term in each IP-Adapter cross-attention block; two insets show the per-location products.}
  \caption{Weight spatialization overview.
  (a)~A content map $s(\mathbf{p})$ scales the ControlNet control embedding (Eq.~\ref{eq:fusion}).
  (b)~A style map $\lambda_b(\mathbf{p})$ scales the decoupled image-attention term in each IP-Adapter block (Eq.~\ref{eq:ipa}).
  Photo by D.~Perevoshchikov on Unsplash.}
  \label{fig:method}
\end{figure*}

Image stylization is the process of transforming an image to mimic a specific artistic medium or art style.
At its core it negotiates the tension between preserving an image's underlying structure and re-rendering its appearance as artistic marks such as brush strokes~\cite{kyprianidis2013stateart}.
Latent-diffusion models (LDMs) redefined this workflow by moving generation into a compact, perceptually compressed latent space, bringing unprecedented visual quality within reach of consumer hardware~\cite{rombach2022ldm}.
However, whereas algorithmic filters transform an image locally, LDMs resynthesize it globally, so neither structure nor appearance stays pinned to the input pixels, turning stylization into a dual-axis process that manipulates \emph{what} a region depicts (content) and \emph{how} it is rendered (style) across the whole image.
Yet professional workflows such as graphic design demand local, spatial control: they realize deliberate design choices region by region, not as one uniform look.
Ideally, both axes should thus be controllable locally.

Parameter-space painting brought exactly this spatial control to heuristic-based stylization, letting the user control where and how strongly an effect applies~\cite{decarlo2002stylization, semmo2016multilevel}.
Conditioning and adapter methods that followed, e.g.,~ControlNet~\cite{zhang2023controlnet} and IP-Adapter~\cite{ye2023ipadapter}, restored some structural and stylistic control.
Yet a ControlNet conditioning image only specifies what structure to follow per location, while how strongly it is enforced and how strongly the image-prompt style is applied remain uniform over the image.

In this work, we bring the local control of parameter-space painting to the generative setting, spanning content as well as style.
We observe that the necessary controls already exist in a standard ControlNet\,+\,IP-Adapter pipeline as two scalar conditioning weights: a ControlNet strength governs adherence to content structure, and a per-block IP-Adapter scale governs adherence to the image-prompt style.
Our contribution is to lift each to a per-location spatial map and to validate quantitatively that the two controls, acting on disjoint pathways, stay local and axis-specific despite global attention mixing.
Together they span a $2{\times}2$ vocabulary of retouching operations, from free regeneration to identity preservation (Fig.~\ref{fig:teaser}).


\section{Related Work}
\label{sec:related}

\paragraph{Spatial diffusion control.}
Paint-with-words~\cite{balaji2022ediffi} and MultiDiffusion~\cite{bar2023multidiffusion} steer the text pathway spatially, orthogonal to the content and style pathways we control.
A broad literature on layout, grounding, and region conditioning (surveyed in~\cite{cao2025controllable}) expands structural control but spatializes \emph{what} is conditioned, not \emph{how strongly} a pathway applies per location.
Spatially varying the IP-Adapter scale is an established practitioner technique, but for style alone.
SmartControl~\cite{liu2024smartcontrol} spatializes the ControlNet scale with a learned predictor that only relaxes a single structural condition where it conflicts with the prompt.
None couples a spatial content strength with a spatial style scale, which is our contribution rather than the regional idea itself.

\paragraph{Image-prompt stylization.}
Adapter-based stylization conditions the generation on a style image rather than text: IP-Adapter~\cite{ye2023ipadapter} injects image-prompt features through decoupled cross-attention, and InstantStyle~\cite{wang2024instantstyle} shows that restricting this injection to specific attention blocks already separates style from content. B-LoRA~\cite{frenkel2024blora} independently confirms that these SDXL~\cite{podell2023sdxl} blocks specialize in style versus content.
Our pipeline builds on this image-prompt configuration, and preserves content through a ControlNet~\cite{zhang2023controlnet} Tile branch in the manner introduced for stylization by InstantStyle-Plus~\cite{wang2024instantstyleplus}.
A broader line separates the two factors more explicitly but \emph{globally}, fixed for the whole forward pass~\cite{qi2024deadiff,xing2024csgo,hertz2024stylealigned,jiang2025diffartist}.

\paragraph{Diffusion image editing.}
Pretrained diffusion models also support stroke-, mask-, and instruction-guided editing~\cite{meng2022sdedit, hertz2023prompt2prompt, avrahami2023blendedlatent, huang2025editsurvey}, but these target \emph{content} (e.g., layout, attributes) and expose no per-location control of image-\emph{style} strength.
In contrast, our spatial maps let both content and style strength be controlled per location.

\paragraph{Level-of-abstraction control.}
Non-photorealistic rendering has long varied an image's \emph{level of abstraction} locally, driven by gaze, saliency, or depth~\cite{kyprianidis2013stateart, decarlo2002stylization} or by parameter maps~\cite{semmo2016multilevel}.
The neural era continued this with perceptual-factor control~\cite{gatys2017} and depth- or saliency-guided style transfer~\cite{jing2020review}.
Our maps are the generative counterpart of this abstraction control: they trade content preservation against style transfer per region.

\begin{figure*}[t]
  \centering
  \begin{minipage}[t]{1.9cm}\centering
    \includegraphics[width=1.9cm,height=1.9cm]{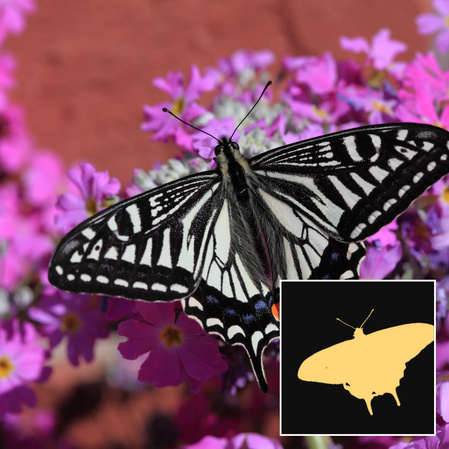}\\
    {\scriptsize source $+$ mask inset\par}
  \end{minipage}\hspace{0.1cm}%
  \begin{minipage}[t]{1.9cm}\centering
    \includegraphics[width=1.9cm,height=1.9cm]{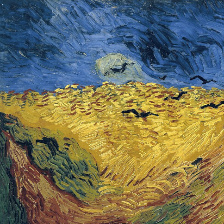}\\
    {\scriptsize style\par}
  \end{minipage}\hspace{0.1cm}%
  \begin{minipage}[t]{2.6cm}\centering
    \includegraphics[width=2.6cm,height=2.6cm]{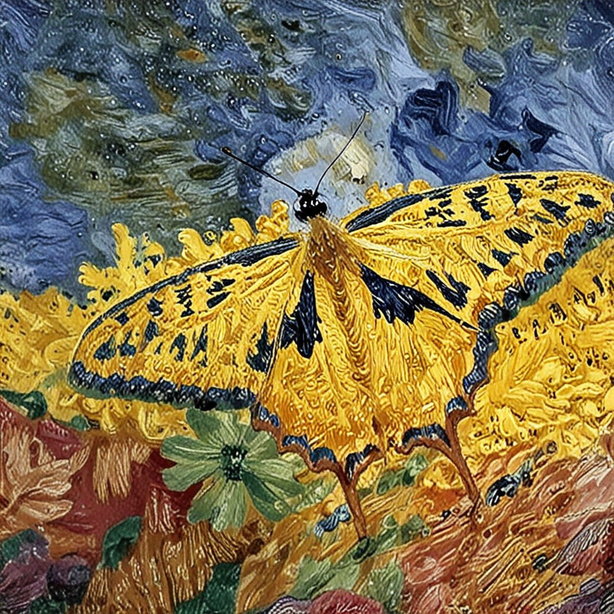}\\
    {\scriptsize $B$ free regen.\par}
  \end{minipage}\hspace{0.1cm}%
  \begin{minipage}[t]{2.6cm}\centering
    \includegraphics[width=2.6cm,height=2.6cm]{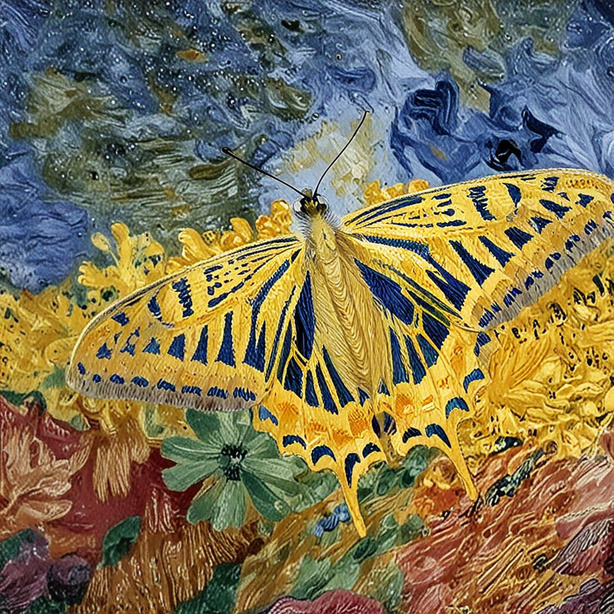}\\
    {\scriptsize $C$ content-lock\par}
  \end{minipage}\hspace{0.1cm}%
  \begin{minipage}[t]{2.6cm}\centering
    \includegraphics[width=2.6cm,height=2.6cm]{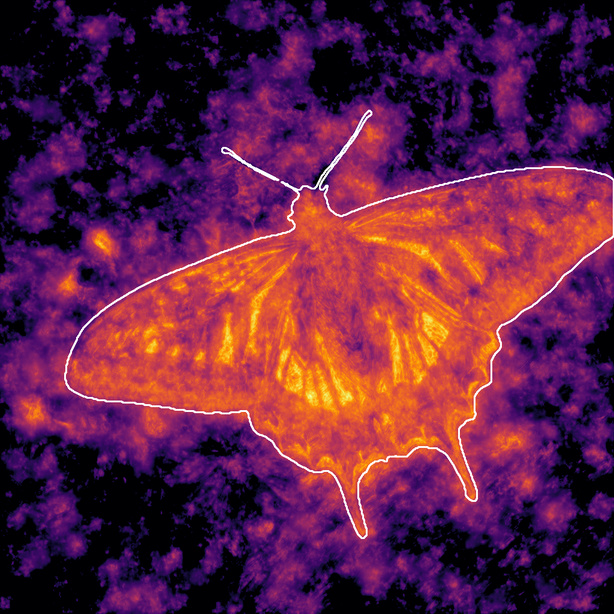}\\
    {\scriptsize $|\Delta|$ $C$ vs.\ $B$\par}
  \end{minipage}\hspace{0.1cm}%
  \begin{minipage}[t]{2.6cm}\centering
    \includegraphics[width=2.6cm,height=2.6cm]{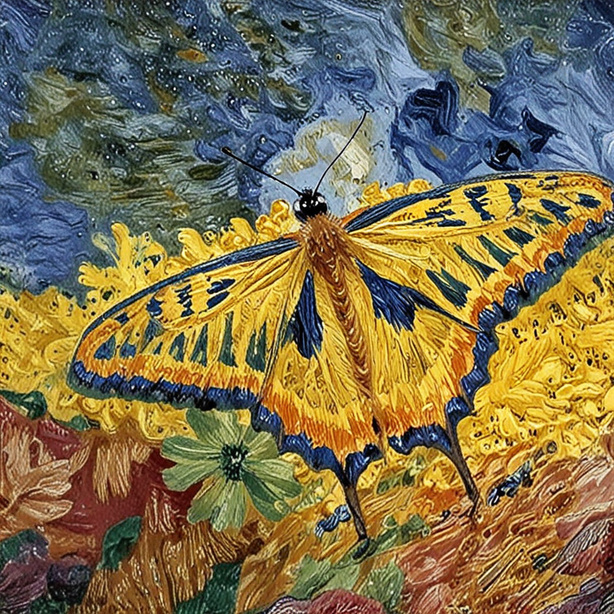}\\
    {\scriptsize $S$ style-relax\par}
  \end{minipage}\hspace{0.1cm}%
  \begin{minipage}[t]{2.6cm}\centering
    \includegraphics[width=2.6cm,height=2.6cm]{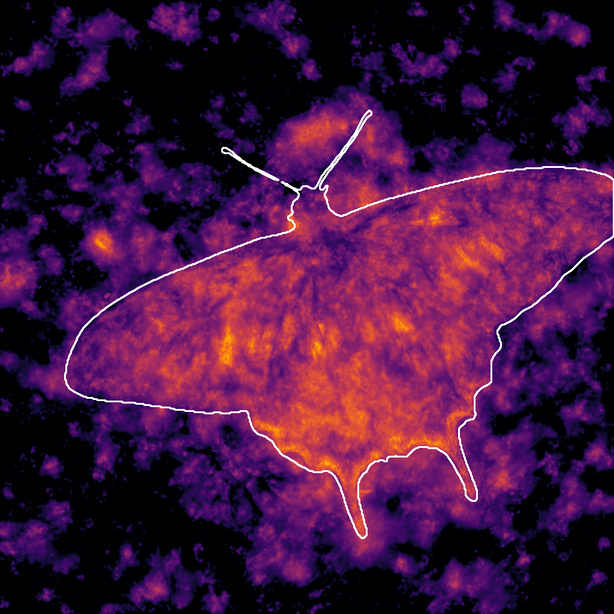}\\
    {\scriptsize $|\Delta|$ $S$ vs.\ $B$\par}
  \end{minipage}
  \Description{A butterfly photo with its mask, a style image, and three results (default stylization, content-lock, style-relax), with heatmaps of change relative to the default that are bright inside the butterfly and dark in the background.}
  \caption{Locality and specificity on one evaluation triple (Table~\ref{tab:disentangle}).
  Content-lock ($C$) recovers the wing markings and style-relax ($S$) restores their contrast, while the change ($|\Delta|$ heatmaps, LPIPS vs.\ baseline $B$) stays confined to the masked butterfly.}
  \label{fig:locviz}
\end{figure*}

\section{Method}
\label{sec:method}

\subsection{Pipeline and Modulation Points}

Our method spatializes the two scalar conditioning weights of a standard latent-diffusion text-to-image pipeline~\cite{rombach2022ldm} (Fig.~\ref{fig:method}), whose denoising U-Net is conditioned on content structure by a ControlNet branch~\cite{zhang2023controlnet} and on image-prompt style by an IP-Adapter path~\cite{ye2023ipadapter}.
Driving the ControlNet with a Tile control turns this pipeline into an image stylizer~\cite{wang2024instantstyleplus}.
The formulation depends only on these two pathways, not on a particular backbone, sampler, or distillation.
Sec.~\ref{sec:ondevice} describes our SDXL-based~\cite{podell2023sdxl} instantiation.

\subsection{Content Axis: Spatial ControlNet}
The ControlNet branch embeds the noisy latent $z_t$ into a feature map $h$ and the control image (Tile, in our pipeline) into a control embedding $c$ at the same latent resolution~\cite{zhang2023controlnet}, combined with a conditioning strength $s$ into the conditioned embedding $\tilde h = h + s\, c$.
We replace the scalar $s$ with a spatial map $s \in \mathbb{R}^{H\times W}$ at latent resolution, broadcast over channels:
\begin{equation}
  \tilde h(\mathbf{p}) = h(\mathbf{p}) + s(\mathbf{p})\, c(\mathbf{p});
  \label{eq:fusion}
\end{equation}
a uniform map reproduces the scalar exactly.

\subsection{Style Axis: Spatial IP-Adapter}
Each IP-Adapter--injected block adds a decoupled image-attention term to the text attention~\cite{ye2023ipadapter}:
\begin{equation}
  Z^{\mathrm{new}} = \mathrm{Attn}(Q,K,V) + \lambda_b\,\mathrm{Attn}(Q,K',V'),
  \label{eq:ipa}
\end{equation}
where the shared query $Q$ comes from the block's features, $(K,V)$ and $(K',V')$ are the text and image-prompt keys and values, and $\lambda_b$ is the per-block image scale ($\lambda_b{=}0$ recovers the base model).
We lift the scalar $\lambda_b$ to a per-location map $\lambda_b(\mathbf{p})$.
In our accelerator-friendly attention layout, the image-attention output is shaped $[B,C,1,HW]$ (one feature vector per latent position), so $\lambda_b(\mathbf{p})$ broadcasts over channels with no change to the attention computation itself.
This way, only the block's scale input changes shape, and a uniform map again reproduces the scalar exactly.

\subsection{Resolution Semantics}
The pixel-resolution maps are anti-alias downsampled to each modulation point.
Content modulation thus propagates multi-scale through the down-blocks---crisp where decisions are fine, soft where they are coarse---while style modulation is inherently coarser.
This asymmetry suits both factors: structure benefits from fine spatial control whereas the spatial assignment of style is a low-frequency, semantic choice.

\section{Results and Evaluation}
\label{sec:eval}

\paragraph{Implementation.}
\label{sec:ondevice}
We instantiate our method in a 4-step-distilled SDXL-Lightning~\cite{lin2024sdxllightning} pipeline with ControlNet and IP-Adapter, converted for the Apple Neural Engine, where the spatial inputs add negligible latency and memory over the scalar pipeline.
At its $1024^2$ output resolution, the maps downsample to latent $128^2$ for ControlNet and $32^2$/$64^2$ for the injected attention blocks.

\paragraph{The $2{\times}2$ retouching vocabulary.}
The two axes span a $2{\times}2$ vocabulary (Fig.~\ref{fig:teaser}): content is \emph{locked} or \emph{relaxed} around a balanced default ($s{=}0.5$), while style starts at full strength ($\lambda_b{=}1.0$) and typically only \emph{relaxes}.
Locking content preserves structure (\emph{structure preserved}), relaxing style pulls the output toward the content image (\emph{stylization weakened}), and both together can preserve a subject's identity amid the stylized scene (\emph{identity preserved}).
Relaxing either axis instead gives the text prompt more weight in the region, e.g., for content swaps (supplemental Fig.~S1).
Because the Tile input is the content image, content-lock also displaces some stylization, which we can offset by co-raising $\lambda_b$ inside the region.

\begin{table}[t]
  \centering
  \caption{Locality (H1) and specificity (H2) over $N{=}300$ triples ($10$ contents $\times$ $10$ styles $\times$ $3$ seeds). \emph{H1}: $\ell_\infty$ relative to the inside response, stylization magnitude, and a seed re-roll (lower $=$ more local). \emph{H2}: signed responses (negative $=$ toward the source; compare within columns---probe scales differ); content loads on structure (edges), style on texture.}
  \label{tab:disentangle}
  \small
  \resizebox{\columnwidth}{!}{%
  \begin{tabular}{@{}l ccc ccc@{}}
    \toprule
    & \multicolumn{3}{c}{H1: $\ell_\infty$ relative to ($\downarrow$)} & \multicolumn{3}{c}{H2: signed probe response} \\
    \cmidrule(lr){2-4}\cmidrule(lr){5-7}
    Control & inside & styliz. & roll & edges & texture & palette \\
    \midrule
    lock content ($s{\to}0.8$)        & 0.33 & 0.18 & 0.24 & $-0.026$ & $-0.046$ & $-0.003$ \\
    relax style ($\lambda_b{\to}0.7$) & 0.26 & 0.10 & 0.13 & $-0.004$ & $-0.024$ & $+0.003$ \\
    relax style ($\lambda_b{\to}0.4$) & 0.27 & 0.16 & 0.21 & $-0.006$ & $-0.063$ & $-0.005$ \\
    \bottomrule
  \end{tabular}}
\end{table}

\begin{figure*}[t]
  \centering
  \begin{minipage}[t]{5.75cm}\centering
    \includegraphics[width=1.85cm,height=1.85cm]{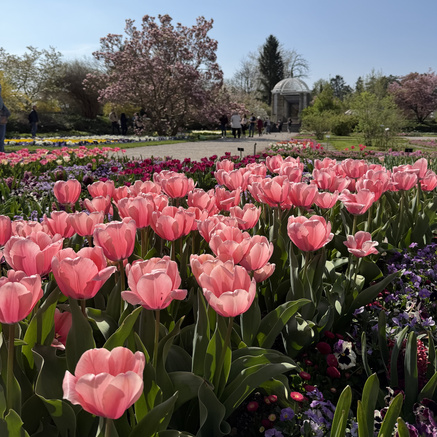}\hspace{0.1cm}\includegraphics[width=1.85cm,height=1.85cm]{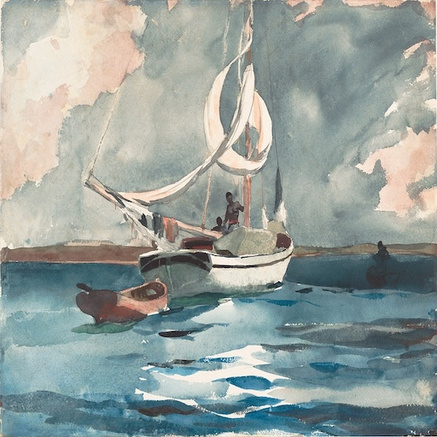}\hspace{0.1cm}\includegraphics[width=1.85cm,height=1.85cm]{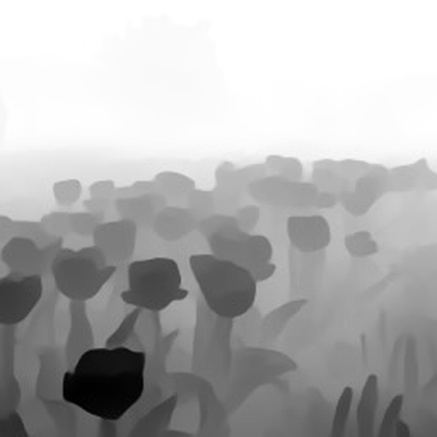}\\[0.25ex]
    \includegraphics[width=2.8cm,height=2.8cm]{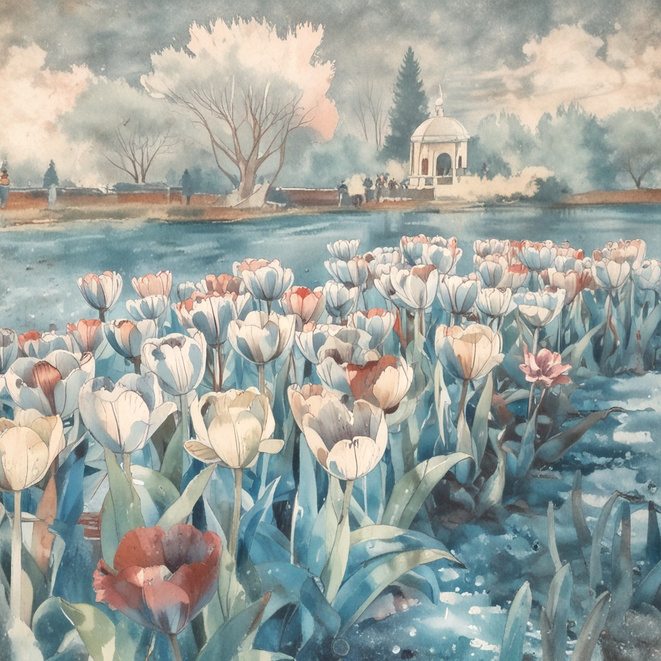}\hspace{0.1cm}\includegraphics[width=2.8cm,height=2.8cm]{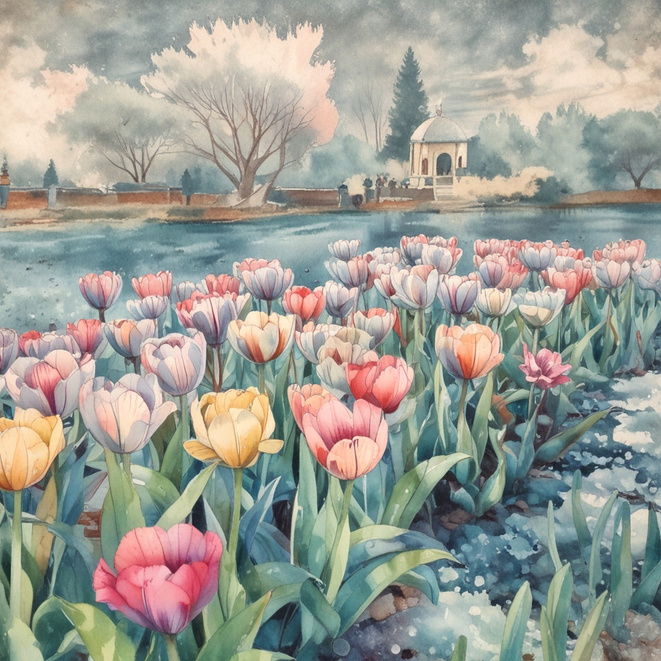}\\
    {\scriptsize depth $\rightarrow$ aerial perspective\par}
    {\tiny prompt: ``watercolor style''\par}
    {\tiny style-block $\lambda_6(\mathbf{p}) = \mathrm{depth}(\mathbf{p})$: exemplar off up close\par}
  \end{minipage}\hspace{0.1cm}%
  \begin{minipage}[t]{5.75cm}\centering
    \includegraphics[width=1.85cm,height=1.85cm]{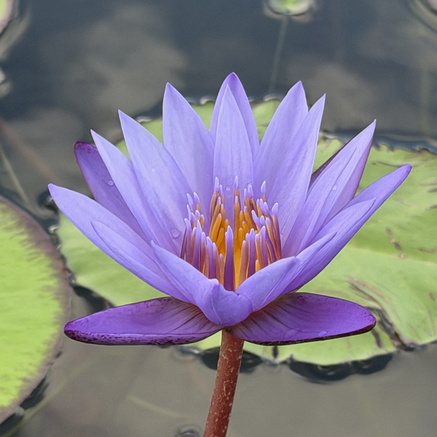}\hspace{0.1cm}\includegraphics[width=1.85cm,height=1.85cm]{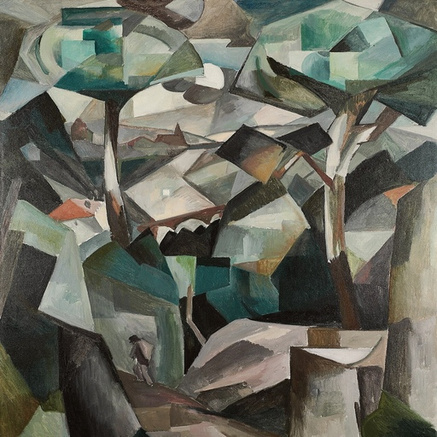}\hspace{0.1cm}\includegraphics[width=1.85cm,height=1.85cm]{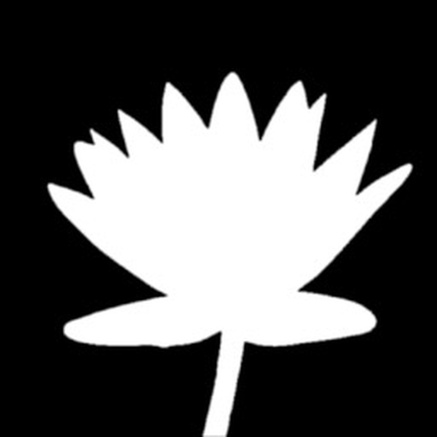}\\[0.25ex]
    \includegraphics[width=2.8cm,height=2.8cm]{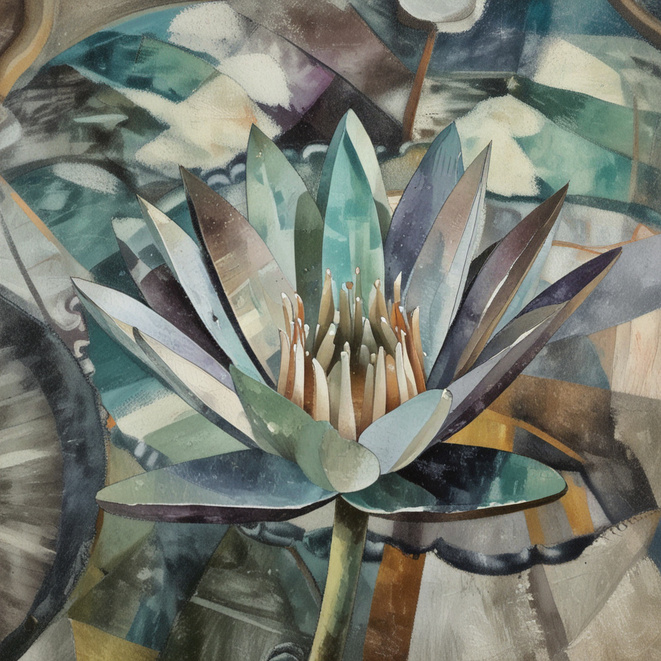}\hspace{0.1cm}\includegraphics[width=2.8cm,height=2.8cm]{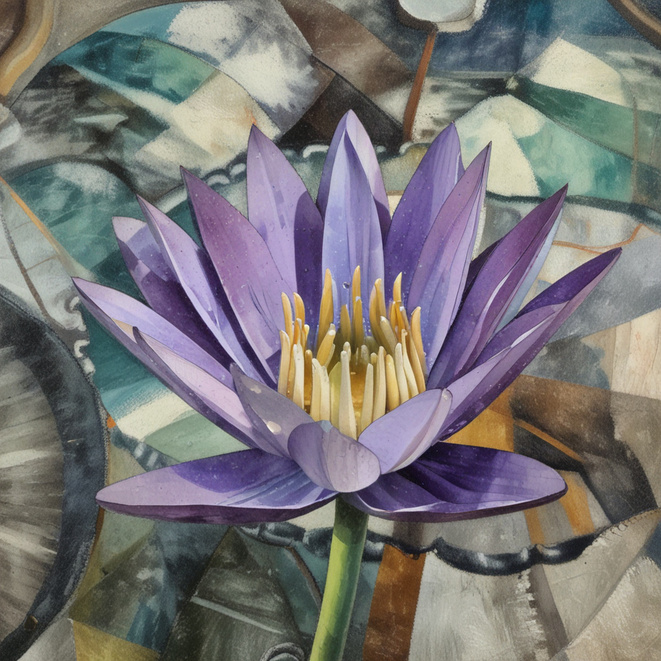}\\
    {\scriptsize subject $\rightarrow$ prompt color unblocked\par}
    {\tiny prompt: ``\textbf{purple} water lily, cubist painting, \dots''\par}
    {\tiny style-block $\lambda_6$: $1.0 \rightarrow 0.3$ on the lily\par}
  \end{minipage}\hspace{0.1cm}%
  \begin{minipage}[t]{5.75cm}\centering
    \includegraphics[width=1.85cm,height=1.85cm]{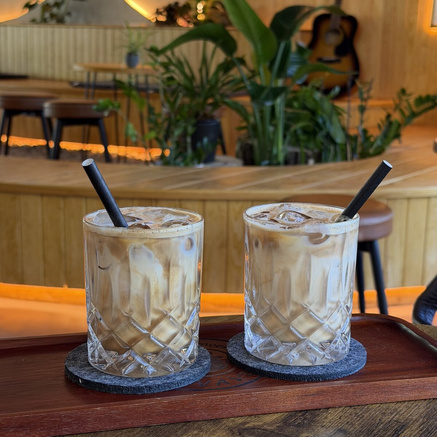}\hspace{0.1cm}\includegraphics[width=1.85cm,height=1.85cm]{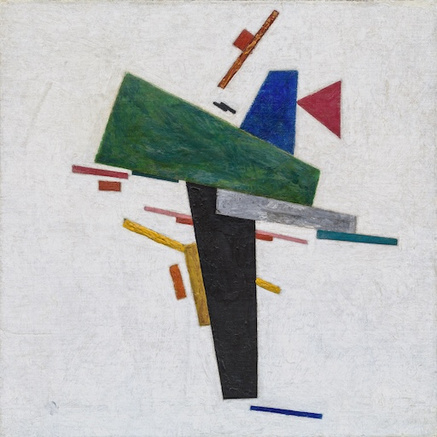}\hspace{0.1cm}\includegraphics[width=1.85cm,height=1.85cm]{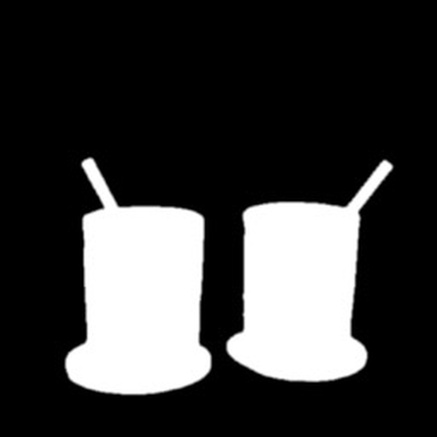}\\[0.25ex]
    \includegraphics[width=2.8cm,height=2.8cm]{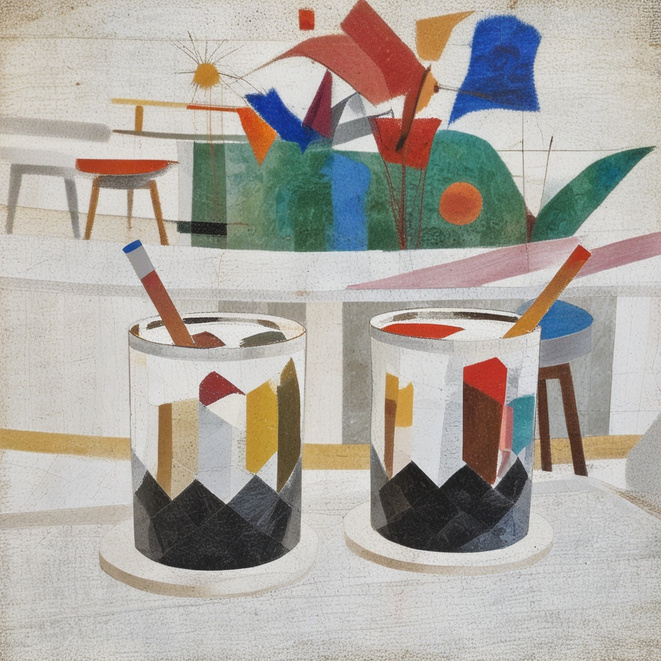}\hspace{0.1cm}\includegraphics[width=2.8cm,height=2.8cm]{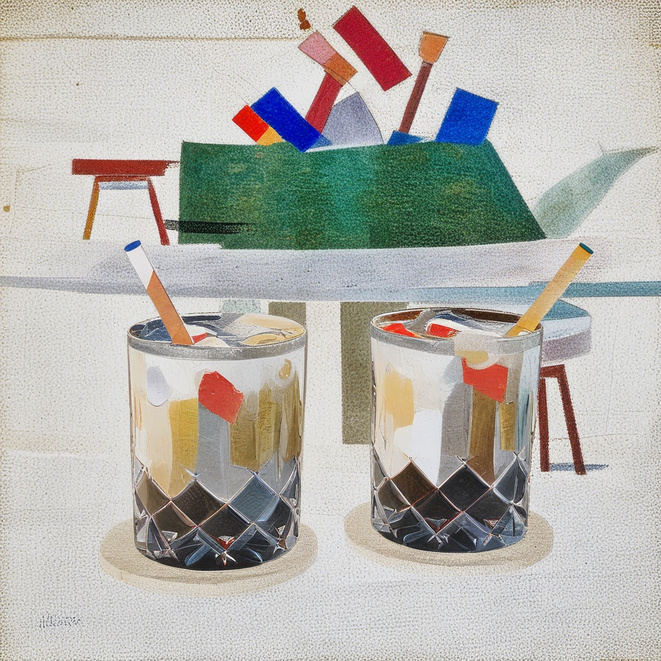}\\
    {\scriptsize objects $\rightarrow$ stylized focus pull\par}
    {\tiny prompt: ``suprematist composition, \dots''\par}
    {\tiny outside: all-block style; inside: $s$: $0.5 \rightarrow 0.9$, style trimmed\par}
  \end{minipage}
  \Description{Three demos, each with a source photo, a style image, and an automatically derived control mask above a uniform and a map-driven result: depth controlling watercolor strength over a tulip field, a subject mask freeing a water lily's purple color, and an object mask sharpening two glasses in a suprematist scene.}
  \caption{Control maps from perceptual priors: source, style, and mask (top) and uniform vs.\ map-driven result (bottom).
  Depth sets the style weight (left), relaxing style frees the prompt's ``purple'' (middle), and an object mask sharpens the glasses (right).}
  \label{fig:usecases-priors}
\end{figure*}

\paragraph{Evaluation.}
We evaluate two claims over (content, style, seed) triples (objects and scenes from DIV2K~\cite{agustsson2017div2k}).
All conditions share the same noise, so every difference is attributable to the maps, and all $11$ IP-Adapter blocks carry style (the worst case for confinement).
\emph{H1 (locality)}: an edit confined to region $R$ should change the output predominantly inside $R$.
We measure the far-field change $\ell_\infty$, the mean spatial LPIPS vs.\ the same-seed baseline over pixels beyond $0.25$ image widths from $R$.
Despite self-attention and residual coupling, it stays well below the inside-$R$ response, the stylization magnitude, and even a seed re-roll ($0.13$--$0.24\times$; Table~\ref{tab:disentangle}, top).
\emph{H2 (specificity)}: probing structure by tolerant edge agreement and stylization by VGG-Gram texture and Lab-chroma palette distance, all to the source, content-lock is \emph{structure-dominant}, style-relax's texture response strengthens with dose, the palette stays flat, and style-relax leaves structure unchanged or slightly restored (Table~\ref{tab:disentangle}, bottom and Fig.~\ref{fig:locviz}).

\paragraph{Use cases.}
Because the maps are continuous, sweeping the style scale and content strength over a fixed region moves it gradually from photographic to fully stylized (supplemental Fig.~S2).
The maps can also be derived from per-pixel priors such as depth or segmentation (Fig.~\ref{fig:usecases-priors}), echoing the gaze-directed abstraction of NPR~\cite{decarlo2002stylization}.
Neural style transfer supports spatial weighting as well~\cite{gatys2017}, but it can only restyle the given content, whereas our content pathway can also regenerate a region under the prompt.
Practitioner alternatives fall short: global scalars are not local, regional text prompts steer a different pathway, and stylize-then-composite merges two independently denoised results.

\section{Limitations}
\label{sec:limitations}
Style localizes more coarsely than content because SDXL caps style modulation at $64^2$ latent resolution.
The $8\times$ pixel-to-latent downsampling leaves very small objects hard to control, and the $4$-step distilled sampler used in our experiments resolves intermediate strengths only coarsely.
Finally, relaxing style can reintroduce color into near-monochromatic styles (e.g., pen-and-ink), either from the source or through hallucination.

\section{Conclusion}
\label{sec:conclusion}
We spatialize two conditioning weights of a ControlNet\,+\,IP-Adapter stylization pipeline into per-location maps, yielding local, axis-specific control of content and style in a single coherent generation pass, with no retraining.
This brings region-by-region control of professional retouching to diffusion stylization: control maps pin a subject, relax a background, or adjust abstraction, while per-pixel priors author them automatically.
Future work includes additional ControlNet content types, crop-and-refine for small objects, learned mask priors, and spatial multi-style assignment.

\begin{acks}
This work was funded by the Federal Ministry for Economic Affairs and Energy (BMWi), Germany, for the ZIM project 16KN124010.
\end{acks}

\bibliographystyle{ACM-Reference-Format}
\bibliography{references}


\begin{thebibliography}{26}


\ifx \showCODEN    \undefined \def \showCODEN     #1{\unskip}     \fi
\ifx \showISBNx    \undefined \def \showISBNx     #1{\unskip}     \fi
\ifx \showISBNxiii \undefined \def \showISBNxiii  #1{\unskip}     \fi
\ifx \showISSN     \undefined \def \showISSN      #1{\unskip}     \fi
\ifx \showLCCN     \undefined \def \showLCCN      #1{\unskip}     \fi
\ifx \shownote     \undefined \def \shownote      #1{#1}          \fi
\ifx \showarticletitle \undefined \def \showarticletitle #1{#1}   \fi
\ifx \showURL      \undefined \def \showURL       {\relax}        \fi
\providecommand\bibfield[2]{#2}
\providecommand\bibinfo[2]{#2}
\providecommand\natexlab[1]{#1}
\providecommand\showeprint[2][]{arXiv:#2}
\makeatletter
\@ifundefined{NAT@parse@date}{}{\let\NAT@parse@date@orig\NAT@parse@date}
\@ifundefined{NAT@parse@date}{}{\def\NAT@parse@date#1#2#3#4#5#6@@{\NAT@parse@date@orig#1#2#3#4#5#6@@\def\NAT@tempyear{0000}\def\NAT@tempexlab{{?}}\ifx\NAT@year\NAT@tempyear\ifx\NAT@exlab\NAT@tempexlab\def\NAT@date{[n.\,d.]}\else\edef\NAT@date{[n.\,d.]\NAT@exlab}\fi\fi}}
\makeatother

\bibitem[Agustsson and Timofte(2017)]%
        {agustsson2017div2k}
\bibfield{author}{\bibinfo{person}{Eirikur Agustsson} {and}
  \bibinfo{person}{Radu Timofte}.} \bibinfo{year}{2017}\natexlab{}.
\newblock \showarticletitle{{NTIRE} 2017 Challenge on Single Image
  Super-Resolution: Dataset and Study}. In \bibinfo{booktitle}{\emph{Proc.
  IEEE/CVF Conf. on Computer Vision and Pattern Recognition Workshops
  (CVPRW)}}. \bibinfo{pages}{126--135}.
\newblock


\bibitem[Avrahami et~al\mbox{.}(2023)]%
        {avrahami2023blendedlatent}
\bibfield{author}{\bibinfo{person}{Omri Avrahami}, \bibinfo{person}{Ohad
  Fried}, {and} \bibinfo{person}{Dani Lischinski}.}
  \bibinfo{year}{2023}\natexlab{}.
\newblock \showarticletitle{Blended Latent Diffusion}.
\newblock \bibinfo{journal}{\emph{ACM Transactions on Graphics (TOG)}}
  \bibinfo{volume}{42}, \bibinfo{number}{4} (\bibinfo{year}{2023}),
  \bibinfo{pages}{1--11}.
\newblock


\bibitem[Balaji et~al\mbox{.}(2022)]%
        {balaji2022ediffi}
\bibfield{author}{\bibinfo{person}{Yogesh Balaji}, \bibinfo{person}{Seungjun
  Nah}, \bibinfo{person}{Xun Huang}, \bibinfo{person}{Arash Vahdat},
  \bibinfo{person}{Jiaming Song}, \bibinfo{person}{Qinsheng Zhang},
  \bibinfo{person}{Karsten Kreis}, \bibinfo{person}{Miika Aittala},
  \bibinfo{person}{Timo Aila}, \bibinfo{person}{Samuli Laine},
  \bibinfo{person}{Bryan Catanzaro}, \bibinfo{person}{Tero Karras}, {and}
  \bibinfo{person}{Ming-Yu Liu}.} \bibinfo{year}{2022}\natexlab{}.
\newblock \showarticletitle{{eDiff-I}: Text-to-Image Diffusion Models with an
  Ensemble of Expert Denoisers}.
\newblock \bibinfo{journal}{\emph{arXiv preprint arXiv:2211.01324}}
  (\bibinfo{year}{2022}).
\newblock


\bibitem[Bar-Tal et~al\mbox{.}(2023)]%
        {bar2023multidiffusion}
\bibfield{author}{\bibinfo{person}{Omer Bar-Tal}, \bibinfo{person}{Lior Yariv},
  \bibinfo{person}{Yaron Lipman}, {and} \bibinfo{person}{Tali Dekel}.}
  \bibinfo{year}{2023}\natexlab{}.
\newblock \showarticletitle{{MultiDiffusion}: Fusing Diffusion Paths for
  Controlled Image Generation}. In \bibinfo{booktitle}{\emph{Proc. Int. Conf.
  on Machine Learning (ICML)}}. \bibinfo{pages}{1737--1752}.
\newblock


\bibitem[Cao et~al\mbox{.}(2026)]%
        {cao2025controllable}
\bibfield{author}{\bibinfo{person}{Pu Cao}, \bibinfo{person}{Feng Zhou},
  \bibinfo{person}{Qing Song}, {and} \bibinfo{person}{Lu Yang}.}
  \bibinfo{year}{2026}\natexlab{}.
\newblock \showarticletitle{Controllable Generation with Text-to-Image
  Diffusion Models: A Survey}.
\newblock \bibinfo{journal}{\emph{IEEE Transactions on Pattern Analysis and
  Machine Intelligence (TPAMI)}} \bibinfo{volume}{48}, \bibinfo{number}{4}
  (\bibinfo{year}{2026}), \bibinfo{pages}{4771--4791}.
\newblock


\bibitem[DeCarlo and Santella(2002)]%
        {decarlo2002stylization}
\bibfield{author}{\bibinfo{person}{Doug DeCarlo} {and} \bibinfo{person}{Anthony
  Santella}.} \bibinfo{year}{2002}\natexlab{}.
\newblock \showarticletitle{Stylization and Abstraction of Photographs}.
\newblock \bibinfo{journal}{\emph{ACM Transactions on Graphics}}
  \bibinfo{volume}{21}, \bibinfo{number}{3} (\bibinfo{year}{2002}),
  \bibinfo{pages}{769--776}.
\newblock


\bibitem[Frenkel et~al\mbox{.}(2024)]%
        {frenkel2024blora}
\bibfield{author}{\bibinfo{person}{Yarden Frenkel}, \bibinfo{person}{Yael
  Vinker}, \bibinfo{person}{Ariel Shamir}, {and} \bibinfo{person}{Daniel
  Cohen-Or}.} \bibinfo{year}{2024}\natexlab{}.
\newblock \showarticletitle{Implicit Style-Content Separation using {B-LoRA}}.
  In \bibinfo{booktitle}{\emph{Proc. European Conf. on Computer Vision
  (ECCV)}}. \bibinfo{pages}{181--198}.
\newblock


\bibitem[Gatys et~al\mbox{.}(2017)]%
        {gatys2017}
\bibfield{author}{\bibinfo{person}{Leon~A. Gatys},
  \bibinfo{person}{Alexander~S. Ecker}, \bibinfo{person}{Matthias Bethge},
  \bibinfo{person}{Aaron Hertzmann}, {and} \bibinfo{person}{Eli Shechtman}.}
  \bibinfo{year}{2017}\natexlab{}.
\newblock \showarticletitle{Controlling Perceptual Factors in Neural Style
  Transfer}. In \bibinfo{booktitle}{\emph{Proc. IEEE Conf. on Computer Vision
  and Pattern Recognition (CVPR)}}. \bibinfo{pages}{3985--3993}.
\newblock


\bibitem[Hertz et~al\mbox{.}(2023)]%
        {hertz2023prompt2prompt}
\bibfield{author}{\bibinfo{person}{Amir Hertz}, \bibinfo{person}{Ron Mokady},
  \bibinfo{person}{Jay Tenenbaum}, \bibinfo{person}{Kfir Aberman},
  \bibinfo{person}{Yael Pritch}, {and} \bibinfo{person}{Daniel Cohen-Or}.}
  \bibinfo{year}{2023}\natexlab{}.
\newblock \showarticletitle{Prompt-to-Prompt Image Editing with Cross-Attention
  Control}. In \bibinfo{booktitle}{\emph{Proc. Int. Conf. on Learning
  Representations (ICLR)}}.
\newblock


\bibitem[Hertz et~al\mbox{.}(2024)]%
        {hertz2024stylealigned}
\bibfield{author}{\bibinfo{person}{Amir Hertz}, \bibinfo{person}{Andrey
  Voynov}, \bibinfo{person}{Shlomi Fruchter}, {and} \bibinfo{person}{Daniel
  Cohen-Or}.} \bibinfo{year}{2024}\natexlab{}.
\newblock \showarticletitle{Style Aligned Image Generation via Shared
  Attention}. In \bibinfo{booktitle}{\emph{Proc. IEEE/CVF Conf. on Computer
  Vision and Pattern Recognition (CVPR)}}. \bibinfo{pages}{4775--4785}.
\newblock


\bibitem[Huang et~al\mbox{.}(2025)]%
        {huang2025editsurvey}
\bibfield{author}{\bibinfo{person}{Yi Huang}, \bibinfo{person}{Jiancheng
  Huang}, \bibinfo{person}{Yifan Liu}, \bibinfo{person}{Mingfu Yan},
  \bibinfo{person}{Jiaxi Lv}, \bibinfo{person}{Jianzhuang Liu},
  \bibinfo{person}{Wei Xiong}, \bibinfo{person}{He Zhang},
  \bibinfo{person}{Liangliang Cao}, {and} \bibinfo{person}{Shifeng Chen}.}
  \bibinfo{year}{2025}\natexlab{}.
\newblock \showarticletitle{Diffusion Model-Based Image Editing: A Survey}.
\newblock \bibinfo{journal}{\emph{IEEE Transactions on Pattern Analysis and
  Machine Intelligence (TPAMI)}} \bibinfo{volume}{47}, \bibinfo{number}{6}
  (\bibinfo{year}{2025}), \bibinfo{pages}{4409--4437}.
\newblock


\bibitem[Jiang and Chen(2025)]%
        {jiang2025diffartist}
\bibfield{author}{\bibinfo{person}{Ruixiang Jiang} {and}
  \bibinfo{person}{Chang~Wen Chen}.} \bibinfo{year}{2025}\natexlab{}.
\newblock \showarticletitle{{DiffArtist}: Towards Structure and Appearance
  Controllable Image Stylization}. In \bibinfo{booktitle}{\emph{Proc. ACM Int.
  Conf. on Multimedia (ACM MM)}}. \bibinfo{pages}{9598--9607}.
\newblock


\bibitem[Jing et~al\mbox{.}(2020)]%
        {jing2020review}
\bibfield{author}{\bibinfo{person}{Yongcheng Jing}, \bibinfo{person}{Yezhou
  Yang}, \bibinfo{person}{Zunlei Feng}, \bibinfo{person}{Jingwen Ye},
  \bibinfo{person}{Yizhou Yu}, {and} \bibinfo{person}{Mingli Song}.}
  \bibinfo{year}{2020}\natexlab{}.
\newblock \showarticletitle{Neural Style Transfer: A Review}.
\newblock \bibinfo{journal}{\emph{IEEE Transactions on Visualization and
  Computer Graphics (TVCG)}} \bibinfo{volume}{26}, \bibinfo{number}{11}
  (\bibinfo{year}{2020}), \bibinfo{pages}{3365--3385}.
\newblock


\bibitem[Kyprianidis et~al\mbox{.}(2013)]%
        {kyprianidis2013stateart}
\bibfield{author}{\bibinfo{person}{Jan~Eric Kyprianidis}, \bibinfo{person}{John
  Collomosse}, \bibinfo{person}{Tinghuai Wang}, {and} \bibinfo{person}{Tobias
  Isenberg}.} \bibinfo{year}{2013}\natexlab{}.
\newblock \showarticletitle{State of the ``{Art}'': A Taxonomy of Artistic
  Stylization Techniques for Images and Video}.
\newblock \bibinfo{journal}{\emph{IEEE Trans. Vis. Comput. Graph.}}
  \bibinfo{volume}{19}, \bibinfo{number}{5} (\bibinfo{year}{2013}),
  \bibinfo{pages}{866--885}.
\newblock


\bibitem[Lin et~al\mbox{.}(2024)]%
        {lin2024sdxllightning}
\bibfield{author}{\bibinfo{person}{Shanchuan Lin}, \bibinfo{person}{Anran
  Wang}, {and} \bibinfo{person}{Xiao Yang}.} \bibinfo{year}{2024}\natexlab{}.
\newblock \showarticletitle{{SDXL-Lightning}: Progressive Adversarial Diffusion
  Distillation}.
\newblock \bibinfo{journal}{\emph{arXiv preprint arXiv:2402.13929}}
  (\bibinfo{year}{2024}).
\newblock


\bibitem[Liu et~al\mbox{.}(2024)]%
        {liu2024smartcontrol}
\bibfield{author}{\bibinfo{person}{Xiaoyu Liu}, \bibinfo{person}{Yuxiang Wei},
  \bibinfo{person}{Ming Liu}, \bibinfo{person}{Xianhui Lin},
  \bibinfo{person}{Peiran Ren}, \bibinfo{person}{Xuansong Xie}, {and}
  \bibinfo{person}{Wangmeng Zuo}.} \bibinfo{year}{2024}\natexlab{}.
\newblock \showarticletitle{{SmartControl}: Enhancing {ControlNet} for Handling
  Rough Visual Conditions}. In \bibinfo{booktitle}{\emph{Proc. European Conf.
  on Computer Vision (ECCV)}}. \bibinfo{pages}{1--17}.
\newblock


\bibitem[Meng et~al\mbox{.}(2022)]%
        {meng2022sdedit}
\bibfield{author}{\bibinfo{person}{Chenlin Meng}, \bibinfo{person}{Yutong He},
  \bibinfo{person}{Yang Song}, \bibinfo{person}{Jiaming Song},
  \bibinfo{person}{Jun-Yan Zhu}, {and} \bibinfo{person}{Stefano Ermon}.}
  \bibinfo{year}{2022}\natexlab{}.
\newblock \showarticletitle{{SDEdit}: Guided Image Synthesis and Editing with
  Stochastic Differential Equations}. In \bibinfo{booktitle}{\emph{Proc. Int.
  Conf. on Learning Representations (ICLR)}}.
\newblock


\bibitem[Podell et~al\mbox{.}(2024)]%
        {podell2023sdxl}
\bibfield{author}{\bibinfo{person}{Dustin Podell}, \bibinfo{person}{Zion
  English}, \bibinfo{person}{Kyle Lacey}, \bibinfo{person}{Andreas Blattmann},
  \bibinfo{person}{Tim Dockhorn}, \bibinfo{person}{Jonas M{\"u}ller},
  \bibinfo{person}{Joe Penna}, {and} \bibinfo{person}{Robin Rombach}.}
  \bibinfo{year}{2024}\natexlab{}.
\newblock \showarticletitle{{SDXL}: Improving Latent Diffusion Models for
  High-Resolution Image Synthesis}. In \bibinfo{booktitle}{\emph{Proc. Int.
  Conf. on Learning Representations (ICLR)}}. \bibinfo{pages}{1862--1874}.
\newblock


\bibitem[Qi et~al\mbox{.}(2024)]%
        {qi2024deadiff}
\bibfield{author}{\bibinfo{person}{Tianhao Qi}, \bibinfo{person}{Shancheng
  Fang}, \bibinfo{person}{Yanze Wu}, \bibinfo{person}{Hongtao Xie},
  \bibinfo{person}{Jiawei Liu}, \bibinfo{person}{Lang Chen},
  \bibinfo{person}{Qian He}, {and} \bibinfo{person}{Yongdong Zhang}.}
  \bibinfo{year}{2024}\natexlab{}.
\newblock \showarticletitle{{DEADiff}: An Efficient Stylization Diffusion Model
  with Disentangled Representations}. In \bibinfo{booktitle}{\emph{Proc.
  IEEE/CVF Conf. on Computer Vision and Pattern Recognition (CVPR)}}.
\newblock


\bibitem[Rombach et~al\mbox{.}(2022)]%
        {rombach2022ldm}
\bibfield{author}{\bibinfo{person}{Robin Rombach}, \bibinfo{person}{Andreas
  Blattmann}, \bibinfo{person}{Dominik Lorenz}, \bibinfo{person}{Patrick
  Esser}, {and} \bibinfo{person}{Bj{\"o}rn Ommer}.}
  \bibinfo{year}{2022}\natexlab{}.
\newblock \showarticletitle{High-Resolution Image Synthesis with Latent
  Diffusion Models}. In \bibinfo{booktitle}{\emph{Proc. IEEE/CVF Conf. on
  Computer Vision and Pattern Recognition (CVPR)}}.
\newblock


\bibitem[Semmo et~al\mbox{.}(2016)]%
        {semmo2016multilevel}
\bibfield{author}{\bibinfo{person}{Amir Semmo}, \bibinfo{person}{Tobias
  D\"{u}rschmid}, \bibinfo{person}{Matthias Trapp}, \bibinfo{person}{Mandy
  Klingbeil}, \bibinfo{person}{J\"{u}rgen D\"{o}llner}, {and}
  \bibinfo{person}{Sebastian Pasewaldt}.} \bibinfo{year}{2016}\natexlab{}.
\newblock \showarticletitle{Interactive Image Filtering with Multiple
  Levels-of-Control on Mobile Devices}. In \bibinfo{booktitle}{\emph{Proc. ACM
  SIGGRAPH Asia Mobile Graphics and Interactive Applications (MGIA)}}.
  \bibinfo{pages}{2:1--2:8}.
\newblock


\bibitem[Wang et~al\mbox{.}(2024a)]%
        {wang2024instantstyle}
\bibfield{author}{\bibinfo{person}{Haofan Wang}, \bibinfo{person}{Qixun Wang},
  \bibinfo{person}{Xu Bai}, \bibinfo{person}{Zekui Qin}, {and}
  \bibinfo{person}{Anthony Chen}.} \bibinfo{year}{2024}\natexlab{a}.
\newblock \showarticletitle{{InstantStyle}: Free Lunch Towards Style-Preserving
  in Text-to-Image Generation}.
\newblock \bibinfo{journal}{\emph{arXiv preprint arXiv:2404.02733}}
  (\bibinfo{year}{2024}).
\newblock


\bibitem[Wang et~al\mbox{.}(2024b)]%
        {wang2024instantstyleplus}
\bibfield{author}{\bibinfo{person}{Haofan Wang}, \bibinfo{person}{Peng Xing},
  \bibinfo{person}{Renyuan Huang}, \bibinfo{person}{Hao Ai},
  \bibinfo{person}{Qixun Wang}, {and} \bibinfo{person}{Xu Bai}.}
  \bibinfo{year}{2024}\natexlab{b}.
\newblock \showarticletitle{{InstantStyle-Plus}: Style Transfer with
  Content-Preserving in Text-to-Image Generation}.
\newblock \bibinfo{journal}{\emph{arXiv preprint arXiv:2407.00788}}
  (\bibinfo{year}{2024}).
\newblock


\bibitem[Xing et~al\mbox{.}(2024)]%
        {xing2024csgo}
\bibfield{author}{\bibinfo{person}{Peng Xing}, \bibinfo{person}{Haofan Wang},
  \bibinfo{person}{Yanpeng Sun}, \bibinfo{person}{Qixun Wang},
  \bibinfo{person}{Xu Bai}, \bibinfo{person}{Hao Ai}, \bibinfo{person}{Renyuan
  Huang}, {and} \bibinfo{person}{Zechao Li}.} \bibinfo{year}{2024}\natexlab{}.
\newblock \showarticletitle{{CSGO}: Content-Style Composition in Text-to-Image
  Generation}.
\newblock \bibinfo{journal}{\emph{arXiv preprint arXiv:2408.16766}}
  (\bibinfo{year}{2024}).
\newblock


\bibitem[Ye et~al\mbox{.}(2023)]%
        {ye2023ipadapter}
\bibfield{author}{\bibinfo{person}{Hu Ye}, \bibinfo{person}{Jun Zhang},
  \bibinfo{person}{Sibo Liu}, \bibinfo{person}{Xiao Han}, {and}
  \bibinfo{person}{Wei Yang}.} \bibinfo{year}{2023}\natexlab{}.
\newblock \showarticletitle{{IP-Adapter}: Text Compatible Image Prompt Adapter
  for Text-to-Image Diffusion Models}.
\newblock \bibinfo{journal}{\emph{arXiv preprint arXiv:2308.06721}}
  (\bibinfo{year}{2023}).
\newblock


\bibitem[Zhang et~al\mbox{.}(2023)]%
        {zhang2023controlnet}
\bibfield{author}{\bibinfo{person}{Lvmin Zhang}, \bibinfo{person}{Anyi Rao},
  {and} \bibinfo{person}{Maneesh Agrawala}.} \bibinfo{year}{2023}\natexlab{}.
\newblock \showarticletitle{Adding Conditional Control to Text-to-Image
  Diffusion Models}. In \bibinfo{booktitle}{\emph{Proc. IEEE/CVF Int. Conf. on
  Computer Vision (ICCV)}}. \bibinfo{pages}{3836--3847}.
\newblock


\end{thebibliography}

\end{document}